\documentclass[12pt,preprint]{aastex} 
\shorttitle{} 
\shortauthors{} 
 
\begin{document} 
 
\received{} 
\accepted{} 
 
\title{A Mass-accreting Gamma Doradus Pulsator with a Synchronized Core in {\it Kepler} Eclipsing Binary KIC 7385478  }  
 
\author{Zhao Guo$^{1}$ \& Gang Li$^{2}$} 
\affil{${1}$ Center for Exoplanets and Habitable Worlds, Department of Astronomy \& Astrophysics, 525 Davey Laboratory, The Pennsylvania State University, University Park, PA 16802, USA \\
${2}$ Sydney Institute for Astronomy (SIfA), School of Physics, The University of Sydney, NSW 2006, Australia}

\slugcomment{08/08/2019} 

 
\begin{abstract} 

The short-period (P $\approx1.7$ d), Algol-type eclipsing binary KIC 7385478 consists of an F-type primary star ($M_1 \approx 1.71M_{\odot}$) and an evolved K-type secondary ($M_2 \approx 0.37M_{\odot}$) (Ozdarcan \& Ali Dal 2017). We study the variability of the {\it Kepler} light curve and attribute many frequency peaks in the Fourier spectrum to the spot modulation. These frequencies are in the form of orbital harmonics and are highly variable in amplitude. They are most likely from the mass-accreting primary star. In addition, we identify a series prograde dipole g-modes from the primary star which show a quasi-linear period spacing pattern and are very stable in amplitude. The period spacing pattern reveals an asymptotic period spacing value in agreement with fundamental parameters of the primary star and also implies that the near-convective-core rotation rate is almost the same as the orbital period. Thus both the surface and the core of this Gamma Dor pulsator have synchronized with the binary orbit. We find that a lower stellar mass $\approx 1.50 M_{\odot}$ and higher effective temperature are needed in order to be compatible with the asteroseismic constraints from single star evolutionary models.
\end{abstract} 
 
 
 

\section{Introduction}

The surface rotation rates of stars are primarily inferred from the modulation of spots or other activity proxies. Spectroscopy can provide the projected equatorial surface rotational velocities (A-type stars: Zorec \& Royer 2012; B stars: Huang \& Gies 2006) but suffers from the uncertainty of inclination. For some rare slow rotators showing rotational splittings, asteroseismology enables us to measure the envelope-sensitive rotation rates from p-modes and near-core sensitive rotation rates from g-modes. The period spacing patterns of g-mode pulsators such as $\gamma$ Dor and Slowly Pulsating B-stars (SPB) also have made it possible to measure the internal rotation rate near the convective core boundary for a larger number of stars (Van Reeth et al.\ 2016, 2018; Papics et al.\ 2017; Li et al.\ 2019a,b). For main sequence stars, a large sample of stars have been measured and most of them show rigid rotation profiles (Aerts 2019). 


However, this kind of measurements is still lacking for binary stars. Only a handful of main-sequence binary stars have the core and surface rotation rates measured. In the short-period ($P_{orb}=1.22$ days), circularized, and synchronized binary  KIC 9592855 (Guo et al.\ 2017), it is found that both the near-core region and the surface have the same rotation period which is essentially equal to the orbital period. Similarly, Li et al.\ (2019b) reported that the two $\gamma$ Dor pulsators in the short-period  ($P_{orb}< 1$ day) eclipsing binaries KIC 3341457
and KIC 7596250 both show a core-to-surface rotation rate ratio of about one ($0.98$ and $1.00$, respectively). The two F-type hybrid pulsators in the eccentric binary KIC 10080943 ($P_{orb}=15.3$ days) have near-core rotation rates of about 1.06 and 0.74 times of the surface rotation rates (Schmid \& Aerts 2016). In the eccentric triple system HD 201433 (Kallinger et al.\ 2017), 
although the primary star is essentially a solid-body rotator as expected for MS stars, the surface is found to be rotating two orders of magnitude faster than the interior. All these observations highlight the importance of tidal interactions and angular momentum (AM) transfer in binary and multiple systems.


Here we present the asteroseismic analysis for the eclipsing binary KIC 7385478. This binary has been studied in detail by Ozdarcan \& Ali Dal (2017, hereafter OA17). Measurements of fundamental parameters have been performed by combining the {\it Kepler} photometry and ground-based spectroscopy. The primary star is an early-F main sequence star with $T_{\rm eff}=7000 \pm 150 $K. It is also a $\gamma$ Doradus type pulsator with $M_1=1.71\pm 0.08M_{\odot}$ and $R_1=1.59\pm 0.03R_{\odot}$. We confront these measurements from the binary modeling with the information extracted from the variability of the light curves, especially the internal rotation rate and the stellar mass from studying the g-mode pulsations in the primary.

\section{Spot Modulations}                                

Ozdarcan \& Ali Dal (2017) found the light curve modeling of KIC 7385478 can be significantly improved by including a hot spot on the primary. As a semi-detached system, it is likely that this hot spot is the result of mass transfer from the secondary star impacting on the surface of the primary. The evolved K-type secondary can be very active, and is a perfect candidate of chromospherically active variable star. Although it only contributes to less than 15\% of the total flux, we cannot completely rule out the possibility that it contributes to the spot modulations in the observed light curves.

We obtain the {\it Kepler} Simple Aperture Photometry (SAP) light curves of KIC 7385478 from the KASOC\footnote{Kepler Asteroseismic Science Operation Center} website. The light curves are detrended and prepared following procedures outlined in Guo et al.\ (2016). We subtract a re-binned mean in the phase-folded binary light curve and perform a Fourier analysis to the residuals with the Period04 package (Lenz \& Breger 2005). Ideally, we could subtract a better light curve model with spot modulations as that in OA17 before we perform the Fourier analysis. This can suppress the orbital harmonic frequencies in the Fourier spectrum. However, the g modes we used in the analysis are not orbital harmonics and thus not sensitive to this treatment. We present the Fourier amplitude spectrum in the lower panel of Figure 1. Many frequency peaks locate at multiple times of orbital frequency ($N \times f_{orb}=N\times 0.60406$ day$^{-1}$). We have highlighted these peaks with vertical dotted lines. These orbital-harmonic variations are mainly due to the spot modulation and to a much less extent due to imperfect removal of binary light curve. We find these orbital-harmonics peaks are highly variable. This can be seen in the running Fourier spectrum shown in the upper panel. The two frequencies at $f=f_{orb}=0.60406$ day$^{-1}$ and $f=2f_{orb}=1.20812$ day$^{-1}$, being the largest in amplitude, vary by about 2 to 3 parts per thousand. The amplitude modulation will manifest itself as the frequency modulation, producing a bunch of frequency peaks very close to the central $N f_{\rm orb}$ peak. Many orbital harmonics peaks are present, suggesting that the light curve variations from the spot modulation are very non-sinusoidal.

\section{Gravity Modes and the Period Spacing Patterns }

In contrast to the variable frequency peaks at orbital harmonics, there are also many significant frequencies that are very stable. For example, some prominent ones include the two peaks to the right of $f_{orb}$ (0.6563 and 0.7138 day$^{-1}$), and the three peaks between $3f_{orb}$ and $4f_{orb}$ (1.9428, 2.0252, and 2.1178 day$^{-1}$, see Table 1). There is also a stable peak with a high amplitude at slightly lower than $3f_{orb}$ (1.8051 day$^{-1}$). We interpret these frequencies as g-mode pulsations from the primary star. Ozdarcan \& Ali Dal (2017) already found that the primary star is located in the middle of the $\gamma$ Doradus instability strip.
Gravity mode pulsations in $\gamma$ Dor stars are usually very stable \footnote{Non-linear mode coupling can cause frequency, amplitude, and phase variation. However, there is no signature of this effect in this binary.} and are excited by the convective blocking mechanism (Guzik 2000) although turbulent pressure also plays an important role (Xiong et al.\ 2016). These modes have large mode inertia and suffer from the radiative diffusion dissipation. The typical linear damping timescale of g-modes in the observed frequency range of $\gamma$ Dor stars ($1-3$ day$^{-1}$) is quite long, on the order of $\tau \approx 10^6 - 10^8 $ days.

High order g modes (and r modes) in $\gamma$ Doradus stars and Slowly Pulsating B-stars are quasi-equally spaced in pulsation period. The Period (P) vs. Period Spacing ($\Delta P$) diagram has been widely used as a diagnostic tool for asteroseismic inference (Van Reeth et al.\ 2016, Bouabid et al.\ 2013, Ouazzani et al.\ 2017). Following the technique described in Li et al.\ (2019a), we find that a series of g modes near $2$ day$^{-1}$ follows the expected period spacing pattern. The observed $P-\Delta P$ diagram is illustrated in Figure 2. These modes are very likely to be prograde dipole modes for the following reasons. Firstly, the period spacings are too large to be assigned to $l=2$. Secondly, the $P-\Delta P$ has a significant negative slope, indicating prograde modes. Thirdly, sectoral modes ($m=1,-1$) suffer less from the geometric cancelation than the axisymmetric modes ($m=0$) at the observed inclination ($i=70^{\circ}$). Fourthly, all the modes are in the super inertial regime $f > 2f_{rot}=2f_{orb}=1.21$ day$^{-1}$. Finally, observations of a large sample of $\gamma$ Dor show that prograde modes prevail. The identified modes are labeled with blue dotted lines. We use MCMC (emcee, Foreman-Mackey et al.\ 2013) to infer the posteriors of parameters (the near-core rotation frequency $f_{core}$ and the asymptotic period spacing $\Delta\Pi_0=\int{N}d\ln r$, where $N$ is the Brunt frequency). The correlation plot is shown in Figure 3. We derive a near-core rotation rate of $f_{core}=0.64 \pm 0.01$ day$^{-1}$ and an asymptotic period spacing of
 $\Delta\Pi_0= 4230 \pm 70$ seconds (or $\Delta\Pi_{l=1}=\Delta\Pi_0/\sqrt{l(l+1)}= 2991 \pm 49$s).  The value of $f_{core}$ is remarkably close to the orbital frequency ($f_{orb}=0.60406$ day$^{-1}$)\footnote{The difference is not significant here, as we do not take into account the systematic uncertainty of $f_{core}$ in the $P-\Delta P$ inference. Realistic $P-\Delta P$ contains dips arising from trapped modes. This technique is thus only approximate and can only be applied to the flat region.}. As we expect the binary is already synchronized, we thus find evidence that the surface rotation rate is very similar to the core-rotation rate.
Theory predicts that the stellar surface is synchronized first, and the synchronization gradually moves inwards (Goldreich \& Nicholson 1989).
Observational evidence indeed supports this notion, since the dipole g-modes in the SPB pulsator in the binary HD 201433 reveal an accelerated surface and slow interior. For KIC 7385478, the system must be old enough to synchronize the stellar core.

\section{Constraining Stellar Parameters from Asteroseismology }

In this section, we confront the stellar parameters of the primary as derived from the binary modeling in OA17 with the asteroseismic constraints of $\Delta\Pi_{l=1}$. The RV measurements in OA17 have large uncertainties, and thus the derived mass of the primary star is only accurate to about $5\%$ ($M=1.71\pm 0.08 M_{\odot}$). They found the spectra has essentially a solar metallicity and the estimated effective temperature to be $\approx 7000$K.

We construct stellar structure models in the range of $M \in1.5-1.8M_{\odot}$ ($\Delta M=0.1M_{\odot}$) with the MESA evolution code (Paxton et al.\ 2011, 2013, 2015). We adopt the default convention of convection (B{\"o}hm-Vitense 1958) and the convective core overshoot is described by the exponentially decaying prescription (Herwig 2000). We adopt the default solar mixtures `gs98' (Grevesse \& Sauval 1998). We consider models with solar and slight sub-solar metallicity $Z=0.02$ and $0.015$ and consider models including a certain amount of convective core overshooting described by the exponential overshooting parameter $f_{ov}=0.02$.

In Figure 4, we show the evolution of the dipole mode asymptotic period spacing $\Delta\Pi_{l=1}$ for these models from the zero age main sequence  (ZAMS) to the end of main sequence. The general trend is a decreasing of $\Delta\Pi_{l=1}$ and $T_{\rm eff}$ before the short hook at the end of main sequence. We then show the constraints from all the measurements from the effective temperature, mass, radius and asymptotic period spacing: 

$T_{\rm eff}=7000 \pm 300K$ ($\pm 2\sigma$) (black vertical lines), 

$M=1.71\pm 0.16 M_{\odot}$ ($\pm 2\sigma$) (tracks in green, black, blue, red for $1.5, 1.6, 1.7, 1.8M_{\odot}$, respectively), 

$R=1.59\pm 0.06 R_{\odot}$ ($\pm 2\sigma$) (tracks highlighted in bold)

$\Delta\Pi_{l=1}=2991\pm 196$ seconds ($\pm 4\sigma$) (horizontal lines)\footnote{We deliberately adopt a large error box for the period spacing since the systematic uncertainty of the P vs. $\Delta$P diagnostic is difficult to account for.}

Note that all the models satisfying the radius constraints are close to ZAMS. This is expected, as the primary star is likely to be accreting mass from the secondary. It is rejuvenated and should look like a newborn ZAMS star.
And for models to satisfy the  $\pm 2 \sigma$ constraints of $T_{\rm eff}$, the masses have to be lower than 1.6$M_{\odot}$, and preferably as low as 1.5$M_{\odot}$. The models with M=1.5$M_{\odot}$ satisfying all the constraints all have stellar ages less than 0.6 Gyr, and are close to ZAMS. We cannot take these ages as the age of the binary system, but it indeed agrees with the `rejuvenation' state of the primary.

Models satisfying all the constraints have masses about $1.50 \pm 0.05 M_{\odot}$. 
And models with 1.7 and 1.8$M_{\odot}$ cannot satisfy all the constraints. In the spectroscopic analysis of  OA17, the authors infer the $T_{\rm eff}$ by comparing the composite spectrum with synthetic templates. Although the secondary star only contributes to less than 15\% of the total flux, it does contaminate the spectra and thus the inferred $T_{\rm eff}$ is probably slightly underestimated. 

When including the convective core overshooting, the tracks are extended and shifted to the right. This can be seen in the dashed tracks in Figure 4. The $\Delta\Pi_{l=1}$ also slightly increases as the g-mode propagation cavity has a smaller size, and $\Delta\Pi_{l=1}$ is inversely proportional to an integral in this cavity. The increased $\Delta\Pi_{l=1}$ of models with overshooting actually make it harder to satisfy all the observational constraints. The primary effect of decreasing metallicity from $Z=0.02$ to $Z=0.015$ is (lower panel of Figure 4) a shift of the tracks to the high $T_{\rm eff}$ region, since the resulting stellar opacity is also decreased. Decreasing Z also lowers $\Delta\Pi_{l=1}$ slightly, and thus  it makes only models with $M=1.50M_{\odot}$ satisfy all the constraints, and they all locate near the higher limit of the $T_{\rm eff}$ error box. 

We thus conclude that, within the $\pm 2\sigma$ error box of the measured effective temperature, mass, and radius,  the stellar mass has to be lowered to $\approx 1.5M_{\odot}$ to satisfy the asteroseismic constraints. Adding convective-core overshooting and slightly decreasing the metallicity do not improve the fit and but make the best-fitting models slightly hotter. 

The above results are based on single star evolutionary models. This is a caveat of our analysis and we address this issue in the discussion seciton below.

\section{Discussion }

Observations suggest single main sequence $\gamma$ Dor stars and SPB stars rotate uniformly and their rotation rates depend on the angular momentum at birth since AM conservation is expected (Aerts et al.\ 2019). It is also found that a convective overshooting of $f_{ov}=0.005-0.015$\footnote{corresponding to an overshooting extent of $0.05-0.15$ pressure scale height.} near the convective core boundary and some additional diffusive mixing in the envelope is needed to fit observations (Moravveji et al.\ 2015, 2016; Mombarg et al.\ 2019).

In binaries, the stellar rotation rate also depends on the binary orbital evolution. The four short-period ($P_{orb} \approx 1$ day) eclipsing binaries (EBs) KIC 9592855, KIC 3341457, KIC 7596250, and KIC 7385478 all have circular orbits and a surface-to-core rotation ratio of one. These systems must have gone through sufficient AM transfer and have synchronized the stellar interior. Recently, in addition to the known $\gamma$ Dor EBs in literature (Kurtz et al.\ 2015; {\c C}ak{\i}rl{\i} 2015; {\c C}ak{\i}rl{\i} et al.\ 2017; Guo et al.\ 2016, 2017; He{\l}miniak et al.\ 2017a, b; Lee 2016; Lee \& Park 2018; Zhang et al.\ 2018; Li et al.\ 2019b), Gaulme \& Guzik (2019) did a systematic search for pulsating EBs and reported 119 $\gamma$ Dor in EBs, and these are promising targets for searching for period spacing patterns and measuring internal rotation rates and near-core mixings. 
Such measurements to a large sample of binaries with different orbital periods, eccentricities and evolutionary stages will enable us to calibrate the timescale of tidal circularization/synchronization and the angular momentum transfer inside stars.

 Both $\delta$ Scuti and $\gamma$ Dor type pulsations can exist in mass-accreting A- and F-type stars. For example, AS Eri contains a mass-accreting  $\delta$ Scuti star, pulsating at high frequencies $\approx 65$ day$^{-1}$ (Mkrtichian et al.\ 2004). The mass transfer essentially rejuvenates the star, making it look like to be on the ZAMS. Thus this type of stars pulsates with high-frequency p modes which are close to the asymptotic regime and thus are easier for identifying frequency regularities. The frequency regularity is related to the large frequency separation and very helpful to constrain the mean stellar density (Garcia-Hernandez et al.\ 2009, 2015). For g-modes, these stars will show a $P-\Delta P$ pattern with fewer dips (fewer trapped modes), facilitating the mode identification (Miglio et al.\ 2008; Van Reeth et al.\ 2016). These mass-accreting pulsating stars often reside in eclipsing and spectroscopic binaries for which accurate stellar parameters can be measured. Thus they are excellent targets for precise asteroseismology.

The main conclusion of this paper is that the $\gamma$ Dor pulsator in the close binary KIC 7385478 has a synchronized convective core. Our derived constraints on the mass and temperature of the primary star must be treated with caution since only single star evolution models are used. Ideally, binary star evolution models with mass transfer should be adopted. And a true age can be inferred for this interacting binary system. However, there are still lots of uncertainties in these models, especially for mass-transferring stars. The mass-accreting stars may have different bulk stellar parameters (e.g., higher effective temperatures). The Roche-lobe filling stars may have different asteroseismic properties. For example, tidally trapped p modes and tidally perturbed p modes have been found in the close binary HD74423 and U Gru, respectively (Handler et al.\ 2019; Jonston et al.\ 2019, in prep.). The calibration of tidal synchronization and AM transfer requires detailed modeling of the stellar structures, stellar oscillations, mass transfer, and the binary orbit. The increasing number of measurements of internal rotation rates in binary stars motivates us to embark on an endeavor to refine these processes.


 
\acknowledgments 
We are grateful to the anonymous referee whose suggestions improve the quality of this paper. We thank the {\it Kepler} team for making the data publicly available.

 


\clearpage



\begin{figure} 
\begin{center} 
 {\includegraphics[angle=0,height=12cm]{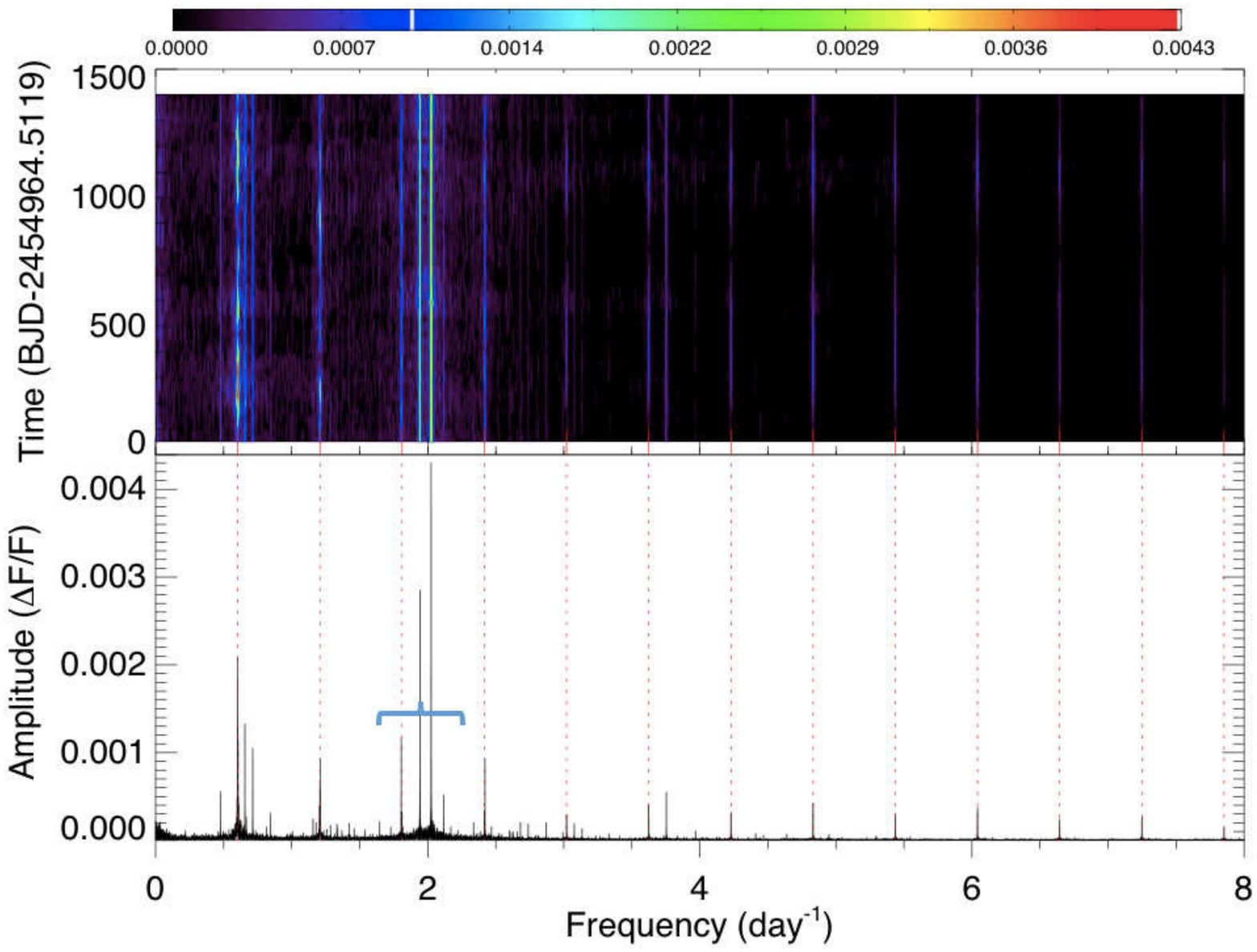}} 
\end{center} 
\caption{\textbf{Lower panel}: The Fourier spectrum of the light curve residuals of KIC7385478. The frequency peaks at orbital harmonics $Nf_{orb}$ are marked by the red dotted lines. \textbf{Upper panel}: A running-window Fourier spectrum showing the variation of pulsation frequencies and their amplitudes. Orbital harmonics frequencies arise from spot modulations and generally show large variations. Other frequency peaks (likely g-modes) are essentially stable over the time span. The blue bracket marks the region where we identified quasi-equally spaced $l=1, m=1$ g modes. }
\end{figure} 
  
\begin{figure} 
\begin{center} 
 {\includegraphics[angle=90,width=18cm,height=14cm]{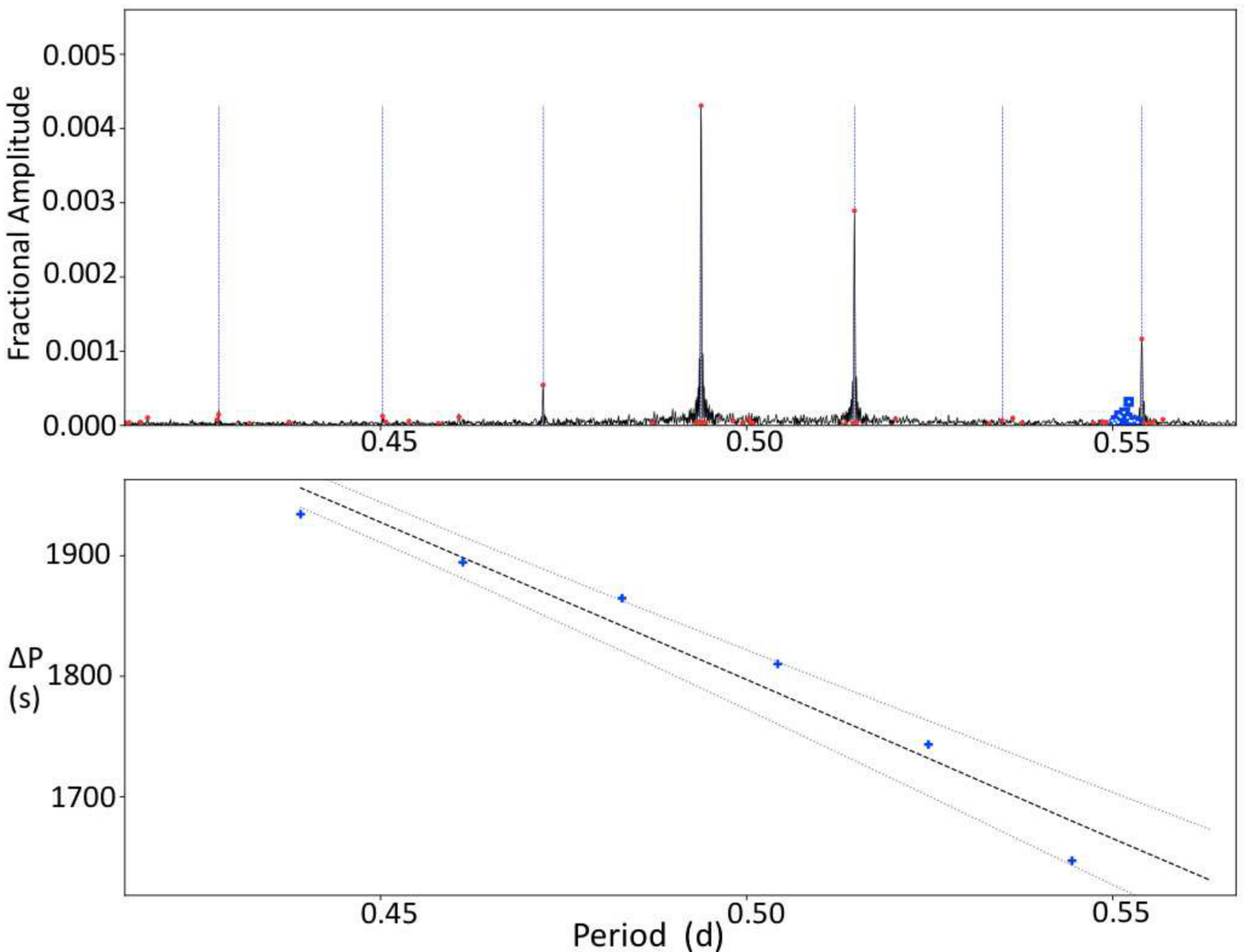}} 
\end{center} 
\caption{\textbf{Upper panel}: The Fourier amplitude spectrum zoomed in to the g modes near $2$ day$^{-1}$ ($P\approx 0.5$ d). These quasi-equally spaced prograde dipole modes are labeled with blue dotted lines. The orbital harmonic near 0.55 d (=$3 \times f_{orb}$) is marked by the blue square. \textbf{Lower panel}:  a fit to the observed P vs. $\Delta$P with the asymptotic period spacing relation. }
\end{figure} 

\begin{figure} 
\begin{center} 
 {\includegraphics[angle=0,height=12cm]{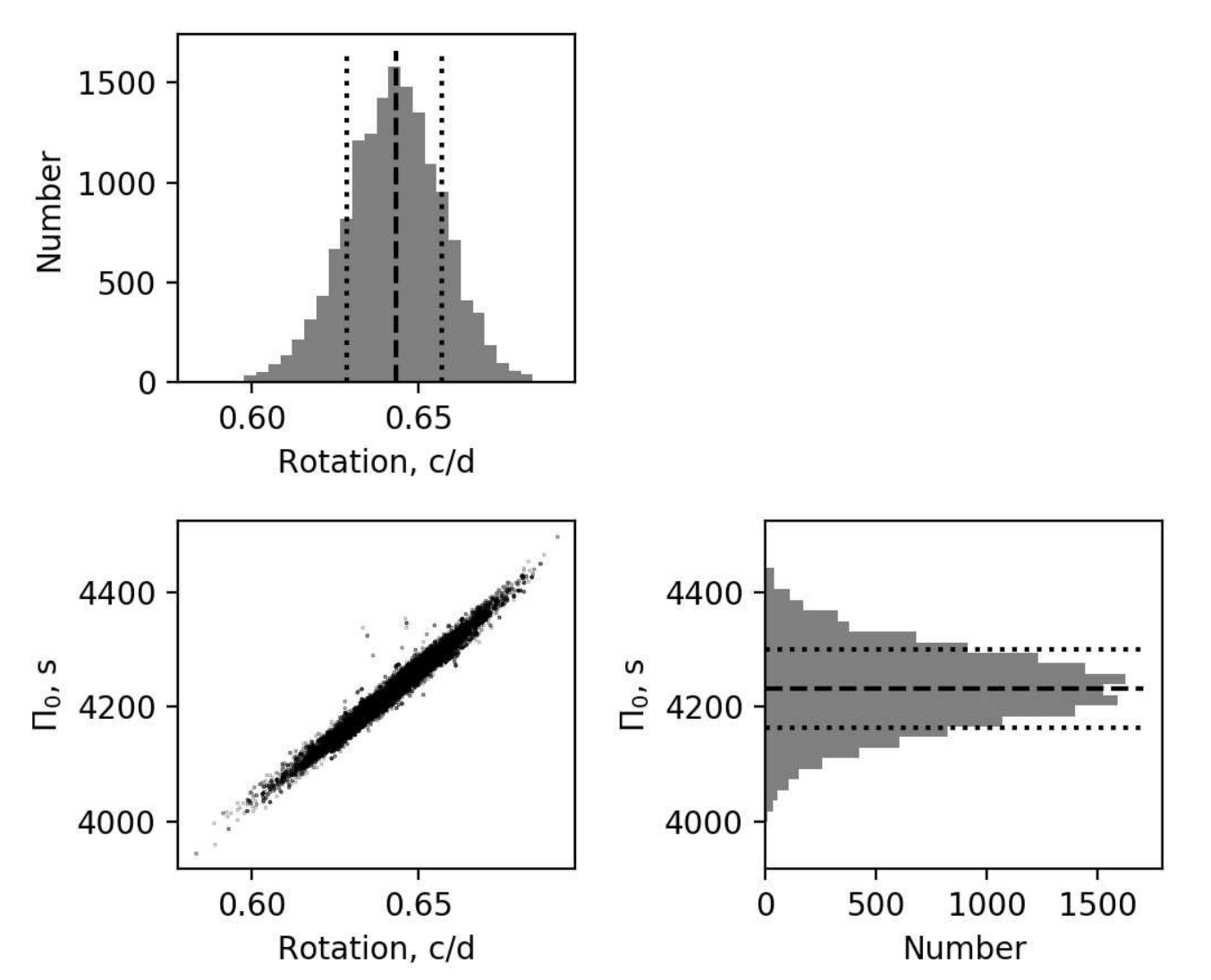}} 
\end{center} 
\caption{Correlation plot of the parameters in the P vs.\ $\Delta$P fitting from the MCMC: asymptotic period spacing $\Delta\Pi_0$ in seconds and the near-core rotation frequency $f_{core}$ (labeled as `Rotation') in cycles per day (c/d). }
\end{figure}

\begin{figure} 
\begin{center} 
 {\includegraphics[angle=0,height=12cm]{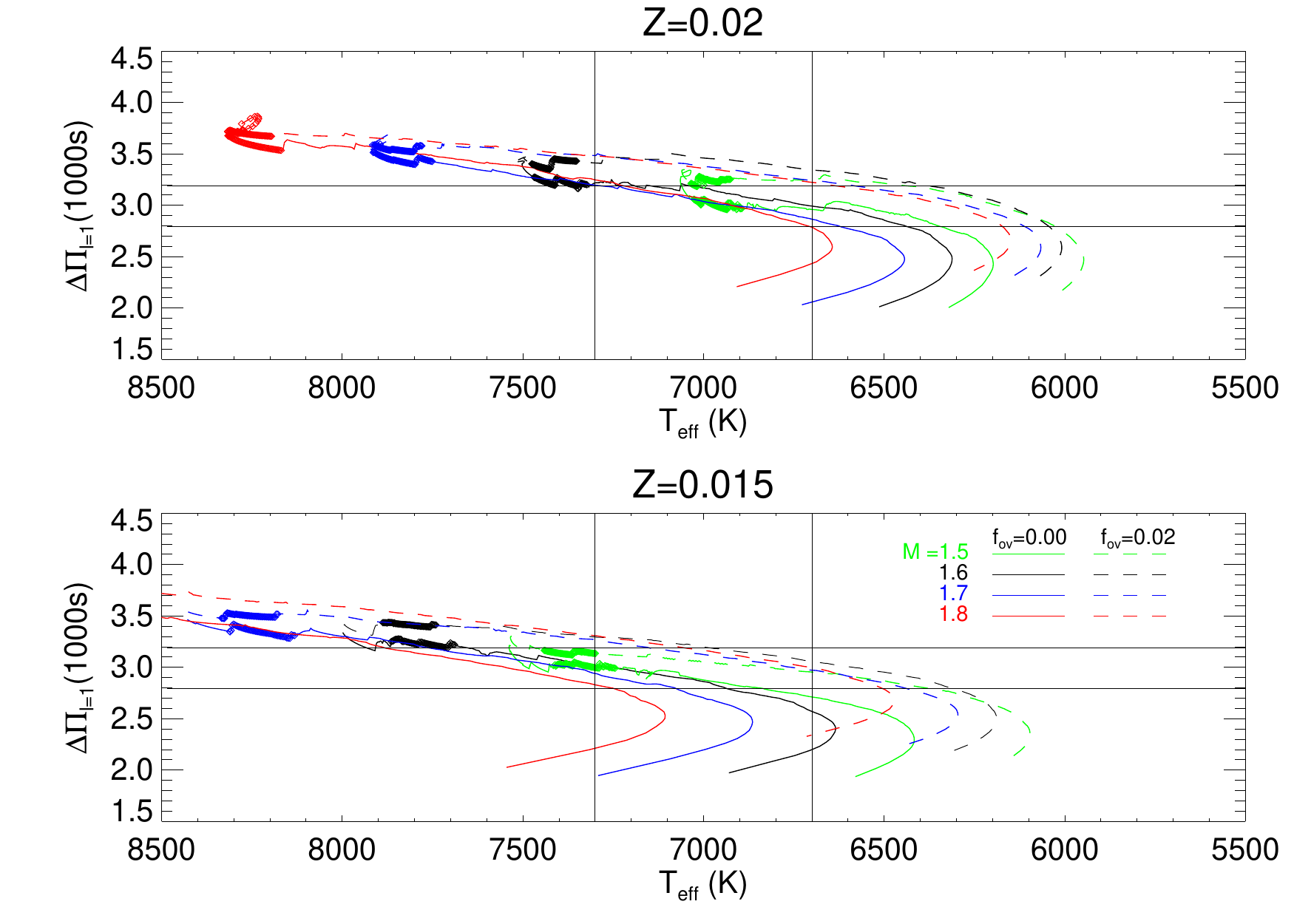}} 
\end{center} 
\caption{\textbf{Upper panel}: The solid tracks show the  evolution of $\Delta\Pi_{l=1}$  as a function of effective temperature $T_{\rm eff}$ from MESA models with masses of $1.5, 1.6, 1.7, 1.8M_{\odot}$ (in green, black, blue, and red, respectively). All  models have solar metallicity (metal mass fraction $Z=0.02$). Those with convective core overshooting $f_{ov}=0.02$ are illustrated by dashed lines. Models on these tracks with stellar radii $R \in 1.59 \pm 0.06 R_{\odot}$ ($\pm 2 \sigma$ region) are over-plotted in diamonds (appearing as bold lines). The vertical and horizontal lines in black show the observational constraints from the $T_{\rm eff} \in 7000 \pm 300$K ($\pm 2 \sigma$ region) and $\Delta\Pi_{l=1} \in 2991 \pm 196$ seconds ($\pm 4 \sigma$ region). \textbf{Lower panel}: same the upper panel but for models with slight sub-solar metallicity $Z=0.015$. }
\end{figure}

 
\begin{deluxetable}{cccccccc}
\tabletypesize{\small} 
\tablewidth{0pc} 
\tablenum{1} 
\tablecaption{Identified g-mode Pulsations\label{tab1}} 
\tablehead{ 
\colhead{Frequency (day$^{-1}$)}          & 
\colhead{Period (days)}          & 
\colhead{Amplitude ($\Delta F/F$)}&
\colhead{Period Spacing (days)}        &     
\colhead{S/N}        &                     
} 
\startdata		
  2.33712(2) &  0.427878(3)     & 0.000147(6)    & 0.022390(4) & 20.0  \\
  2.22090(2) &  0.450267(3)     & 0.000127(5)    & 0.021928(3)  & 18.6\\
  2.117768(4)& 0.4721953(9)    & 0.000544(6)    & 0.0215834(9)  &76.0\\
  2.0251991(7)& 0.4937786(1)    & 0.004308(5)    & 0.0209506(2)  &595.0 \\
  1.9427692(8)& 0.5147292(2)     & 0.002892(6)    & 0.02018(1)  &400.3 \\
  1.86948(4)& 0.53491(1)     & 0.000066(6)    & 0.01906(1)  &8.5 \\
\enddata 
\end{deluxetable}

\end{document}